\documentclass[
reprint,
superscriptaddress,
 amsmath,amssymb,
 aps,
]{revtex4-2}

\usepackage{graphicx}% Include figure files
\usepackage{dcolumn}% Align table columns on decimal point
\usepackage{bm}% bold math

\usepackage{hyperref}% add hypertext capabilities
\usepackage{float} % for forcing figure placements
\usepackage{comment} % 
\usepackage{xcolor}

\makeatletter
\def\maketitle{
\@author@finish
\title@column\titleblock@produce
\suppressfloats[t]}
\makeatother

\begin{document}

\preprint{APS/123-QED}

\title{Transmon qubit using Sn as a junction superconductor
}

\author{Amrita Purkayastha}
\altaffiliation[Current address: ]{Materials Research Laboratory, University of Illinois at Urbana-Champaign, Urbana, IL 61801}
\affiliation{Department of Physics and Astronomy, University of Pittsburgh, Pittsburgh, PA 15260, USA}

\author{Amritesh Sharma}
\affiliation{Department of Physics and Astronomy, University of Pittsburgh, Pittsburgh, PA 15260, USA}

\author{Param J. Patel}
\affiliation{Department of Physics and Astronomy, University of Pittsburgh, Pittsburgh, PA 15260, USA}
\affiliation{Department of Applied Physics, Yale University, New Haven, CT 06511, USA}

\author{An-Hsi Chen}
\affiliation{Univ. Grenoble Alpes, Grenoble INP, CNRS, Institut N\'eel, 38000 Grenoble, France}

\author{Connor P. Dempsey}
\affiliation{Department of Electrical and Computer Engineering,
University of California, Santa Barbara, CA 93106}

\author{Shreyas Asodekar}
\affiliation{Department of Physics and Astronomy, University of Pittsburgh, Pittsburgh, PA 15260, USA}

\author{Subhayan Sinha}
\affiliation{Department of Physics and Astronomy, University of Pittsburgh, Pittsburgh, PA 15260, USA}

\author{Maxime Tomasian}
\altaffiliation[Current address: ]{Univ. Grenoble Alpes, Grenoble INP, CNRS, Institut N\'eel, 38000 Grenoble, France}
\affiliation{Department of Physics and Astronomy, University of Pittsburgh, Pittsburgh, PA 15260, USA}

\author{Mihir Pendharkar}
\altaffiliation[Current address: ]{Department of Materials Science and
Engineering, Stanford University, Stanford, CA 94305}
\affiliation{Department of Electrical and Computer Engineering,
University of California, Santa Barbara, CA 93106}

\author{Christopher J. Palmstr{\o}m}
\affiliation{Department of Electrical and Computer Engineering,
University of California, Santa Barbara, CA 93106}
\affiliation{California NanoSystems Institute, University of California Santa Barbara, Santa Barbara, CA 93106, USA}
\affiliation{Materials Department, University of California Santa Barbara, Santa Barbara, CA 93106, USA}

\author{Moïra Hocevar}
\affiliation{Univ. Grenoble Alpes, Grenoble INP, CNRS, Institut N\'eel, 38000 Grenoble, France}%

\author{Kun Zuo}
\affiliation{School of Physics, The University of Sydney, Sydney, NSW 2006, Australia}
\affiliation{ARC Centre of Excellence for Engineered Quantum Systems, School of Physics,
The University of Sydney, Sydney, NSW 2006, Australia}

\author{Michael Hatridge}
\affiliation{Department of Applied Physics, Yale University, New Haven, CT 06511, USA}

\author{Sergey M. Frolov$^\ast$}
\email{frolovsm@pitt.edu}
\affiliation{Department of Physics and Astronomy, University of Pittsburgh, Pittsburgh, PA 15260, USA}

\date{\today}

\begin{abstract}

Superconductor qubits typically use aluminum-aluminum oxide tunnel junctions to provide the non-linear inductance. Junctions with semiconductor barriers make it possible to vary the superconductor material and explore beyond aluminum. We use InAs semiconductor nanowires coated with thin superconducting shells of $\beta$‑Sn to realize transmon qubits. By tuning the Josephson energy with a gate voltage, we adjust the qubit frequency over a range of 3 GHz. The longest energy relaxation time, $T_1 = 27~\mu$s, is obtained at the lowest qubit frequencies, while the longest echo dephasing time, $T_2 = 1.8~\mu$s, is achieved at higher frequencies. We assess the possible factors limiting coherence times in these devices and discuss steps to enhance performance through improvements in materials fabrication and circuit design.

\end{abstract}

\maketitle

\section{Introduction}

Quantum computing has advanced to the stage where processors with tens-of-qubits have been demonstrated across multiple hardware platforms~\cite{liu2025certified,aghaeerad2025scaling}. Among these approaches, superconducting qubits have emerged as the leading architecture \cite{arute2019quantumsupremacy,google2025QEC} having scaled to the hundreds-of-qubits regime. %But many challenges are yet to be overcome to reach the threshold for fault‑tolerant quantum computation. 
However, many challenges remain before the threshold for fault-tolerant quantum computation can be reached. Leaps forward may come from algorithmic~\cite{google2023suppressing}, architectural~\cite{Putterman2025} and wiring improvements~\cite{krinner2019engineering}. Next to these, advances in materials~\cite{Place2021,bland20252dtransmonslifetimescoherence} for quantum computing stand out because quantum information is extremely sensitive to the microscopic environments both within~\cite{wang2015surface} as well as surrounding the qubits\cite{Houck2008,MicrowavePackage}.

The centerpiece of superconducting qubits is the aluminum-aluminum oxide Josephson junction, a nonlinear electrical element. This is a 50-year old technology in which oxide is typically grown by oxidizing evaporated aluminum, and this will ultimately constrain decoherence times~\cite{McDermott2009,Martinis2005Decoherence,Muller2019TLS}. Research into improving the performance of and adding functionality to Josephson junctions may prove key to scaling up quantum processors~\cite{Kim2021AllNitride,AnferovNiobiumTrilayerQubit2024}. For instance, moving away from magnetic-flux tuning of qubits can save chip real estate and streamline circuitry.

Transmon devices based on semiconductor weak links offer qubit frequency tunability via a gate voltage, providing a faster and more scalable method to tune the qubits. Notable examples include qubits made with Al-InAs nanowires~\cite{larsen2015semiconducting,Luthi2018}, Al-InAs two-dimensional electron gases (2DEGs)~\cite{Casparis2018}, Al-InAs selective area growth (SAG) nanowires~\cite{Hertel2022}, Al-Si/Ge nanowires~\cite{zheng2024coherent,Zhuo2023}, planar Ge~\cite{sagi2024gate} and graphene~\cite{Wang2019}. A wide range of metals can be deposited onto a semiconductor weak link~\cite{pendharkar2021parity,Perla2021NbInAs,Chen2023InSbPb,Bjergfelt2019VInAs, sharma2025sninasnanowireshadowdefinedjosephson}. 
Nevertheless, aluminum remains the dominant choice in hybrid transmons, imposing the requirement of sub-Kelvin temperatures and increasing their susceptibility to pair-breaking radiation ~\cite{Pairbreaking, Serniak2018, Corcoles2011}.

\begin{figure*}[!ht]  
\centering
    \includegraphics[width=15cm]{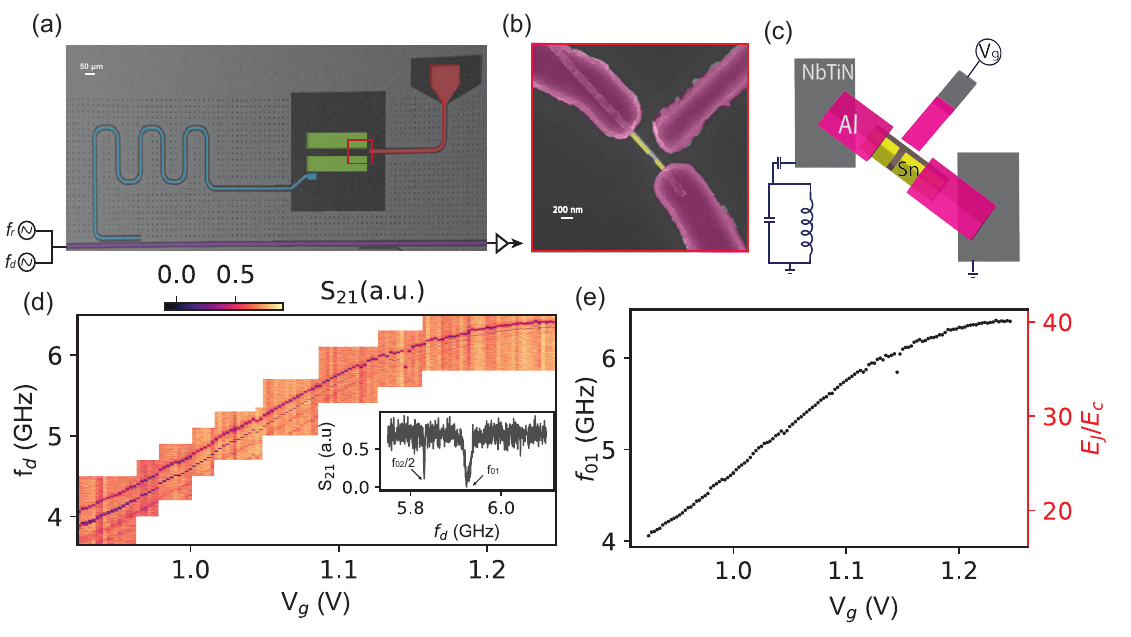}
    \caption{False-colored SEM image of the NbTiN qubit chip. The readout resonator (blue) is inductively coupled to the transmission line (purple) and capacitively coupled to the qubit capacitor (green). The gate port is colored red.
    (b) Schematic of the junction area showing InAs nanowire (dark brown) with Sn shell (yellow) patched with Al (magenta) to the NbTiN ciruit (grey). (c) False-colored SEM image of the Sn-InAs nanowire Josephson junction in the red box in (a). (d) Two-tone spectroscopy showing transmission $S_{21}$ at $f_r$ as a function of qubit drive frequency $f_d$ and $V_g$. Inset: A line cut from (d) at $V_g\text{=1.12~V}$. (e) Extracted \(f_{01}\)(left) and \(E_J/E_c\)(right) as a function of \(V_g\).
}
    \label{fig:Device}
\end{figure*}

We introduce a superconducting transmon qubit based on Sn as a junction superconductor, and InAs as the semiconducting weak link. The superconducting gap of $\beta$-Sn is approximately three times that of Al~\cite{pendharkar2021parity}. 
This can help alleviate the ultralow operating temperature requirements~\cite{AnferovHighTempHighFreq2024}. It may also reduce the impact of quasiparticles on coherence by promoting faster decay dynamics~\cite{QP_lifetimes_Kaplan1976} and enabling their confinement via gap-engineered structures~\cite{GapEngineering2024}. 

InAs is a high-mobility semiconductor~\cite{chuang2013ballistic} that can facilitate gate-voltage tunable Josephson harmonic content~\cite{HarmonicContent_Strambini2020}, introducing functionality not present in oxide tunnel junctions. Gate tunability of our transmon qubit is over 3 GHz which offers additional flexibility in multi-qubit circuits. We attain relaxation times $T_1$ of 27~$\mu s$, likely limited by the substrate and qubit cavity design. Dephasing times $T_2$ are 1.8~$\mu s$, pointing at future improvements such as with preventing partial oxidation of Sn and charging energy optimization.

Identifying superconductor/barrier combinations with long coherence times opens pathways to explore quantum computing beyond aluminum. While our metrics do not overshadow the tunnel junctions at this stage, they show the promise of Sn, and provide a rare example of alternative junctions with coherence times in the tens of microseconds range. Semiconductor InAs nanowires facilitate high transparency barriers to many superconductor metals, making them an attractive template for prototyping materials combinations until a set of the most promising ones is identified. Selective-area grown~\cite{Goswami2023} or top-down etched nanowires~\cite{pendharkar2021parity,sharma2025sninasnanowireshadowdefinedjosephson,Sabbir2023_SnEpitaxy,Sabbir2020_SnTransparentGatable} can be used in the future for scaling up the most promising materials platform.

\section{Device Design}

Scanning electron microscope image of the qubit and the associated superconducting circuitry is shown in Fig.~\ref{fig:Device}(a). The transmon is implemented with a symmetric two-pad floating capacitor where a gate voltage line is routed to the junction area. The qubit is capacitively coupled to a $\lambda/4$ coplanar waveguide (CPW) resonator, with a coupling quality factor of $Q_c \sim 2000$. For the two qubits shown in the main text, the bare resonator frequencies are $f_r=8.182$ GHz for Qubit A and $f_r=7.706$ GHz for Qubit B. The readout resonators are inductively coupled to a common transmission line. The qubits are designed for a charging energy of $E_c/h = e^2 / 2C \approx 380$ MHz.

The Josephson element consists of a semiconducting InAs nanowire coated with a 15 nm thick shell of $\beta$-Sn as shown in Fig.~\ref{fig:Device}(b),(c). The shell covers half of the nanowire circumference. Along the nanowire, the shell is interrupted to create the junction. The break in the shell is achieved by shadowing the junction wire with a nearby wire that acts as a shadow mask. Using a micromanipulator, the nanowire is placed in between the capacitor pads of the pre-fabricated NbTiN circuit indicated by the red box in Fig.~\ref{fig:Device}(a). The nanowire is then connected to the circuit via an Al patch. The schematic of the transmon circuit with the nanowire Josephson junction is shown in Fig.~\ref{fig:Device}(c). The full details of device fabrication can be found in supplementary\ref{App:DevFab}.

\section{Qubit Spectroscopy}

The Josephson energy of superconductor-semiconductor junctions is determined by the gate-tunable critical current, and hence Josephson energy $E_J = \frac{\phi_0 I_c}{2\pi}$, making the qubit frequency $f_{01} = \sqrt{8E_c E_J(V_g)} / h$ also gate tunable. To determine the qubit frequency \(f_{01}\) we perform two-tone spectroscopy applying both the resonator frequency $f_r$ and the drive frequency $f_d$ as a function of gate voltage as shown in Fig.~\ref{fig:Device}(d).

Qubit A frequency shifts monotonically as a function of the gate voltage from $\approx$ 6.5 GHz to $\approx$ 4.5 GHz over a voltage range of \(0.5\,\text{V}\). In separate gate sweeps, we observe \(f_{01}\) as low as $\approx$ 3.5 GHz bringing the total tunable range to 3\,\text{GHz} (see \ref{App:LowFreq}). From the measured \(f_{01}\), we estimate the critical current for Qubit A to be in the range \(I_c \sim 9 - 35\,\text{nA}\). Extracted $f_{01}$ and $E_J/E_c$ as function of $V_g$ are shown in Fig.~\ref{fig:Device}(e). The gate effect is device-specific since InAs nanowire devices exhibit a spread of pinch-off voltages and saturation resistances(see supplementary for \hyperref[QubitB-CW]{QubitB}
).

At relatively high drive power ($P_\mathrm{drive}$), we also observe the two-photon transition from ground-to-second excited state, appearing at $f_{02}/2$, alongside the $f_{01}$ transition, as shown in Fig.~\ref{fig:Device}(d) and highlighted by the linecut in the inset. This allows us to extract the anharmonicity \(\alpha/2h = f_{02}/2 - f_{01}\), which can be considerably reduced from the designed value of charging energy~\cite{KringAnharmonicity2018}. For example, in the inset of Fig.~\ref{fig:Device}(d) the anharmonicity is $\text{192~MHz}$ and it is generally gate-voltage tunable at each $V_g$ which also varies as a function of \(V_g\). This is discussed further in a manuscript currently under preparation.

\section{Rabi Oscillations}

\begin{figure}
\centering
    \includegraphics[width=9 cm]{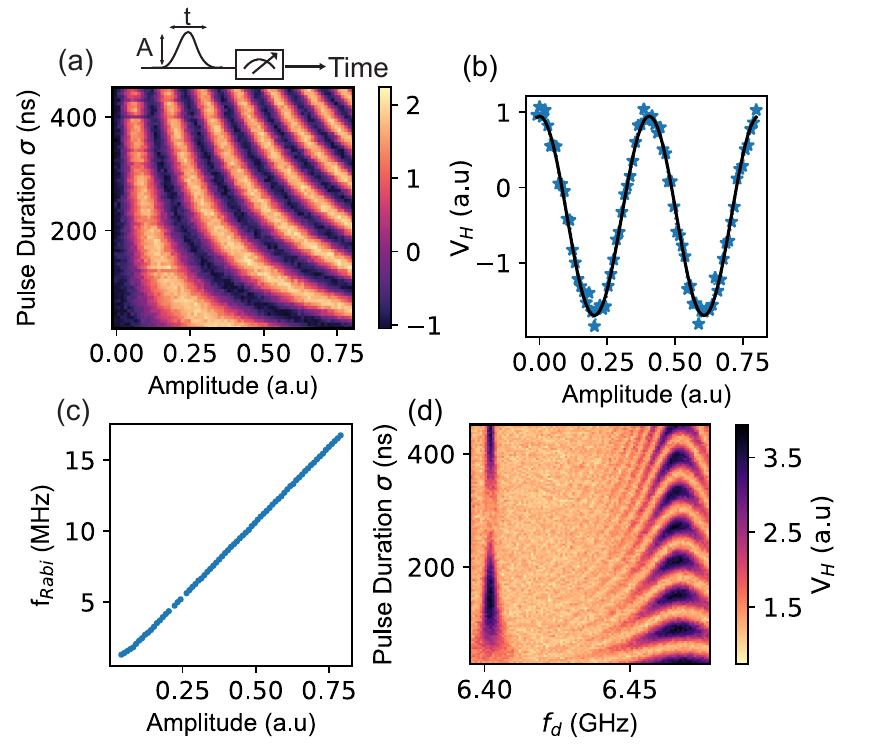}
    \caption{(a) Rabi oscillations as a function of pulse duration and drive amplitude at $f_{01} = \text{6.578~GHz}$ for a side gate voltage of $V_g = 1.38~\text{V}$. (b) A linecut from (a) at pulse duration $\sigma = 120~\text{ns}$. (c) Rabi frequency extracted from (a) as a function of drive. (d) Rabi oscillations as a function of pulse duration and drive frequency at fixed drive amplitude for $V_g = 1.2425~\text{V}$.}

    \label{fig:Rabi Oscillations}
\end{figure}

Next, we demonstrate coherent control of the qubit by inducing Rabi oscillations between its ground and excited states. This is done by applying a pulse at the frequency $f_{01}$ followed by a readout pulse at the resonator frequency (schematic in Fig.~\ref{fig:Rabi Oscillations}(a)). In Fig.~\ref{fig:Rabi Oscillations}(a) we show an example of Rabi oscillations for $f_{01}=\text{6.578 GHz}$. 

A horizontal linecut at $\sigma$=$\text{120 ns}$ shows the oscillations in Fig.~\ref{fig:Rabi Oscillations}(b). As expected for a driven two-level system, the Rabi frequency demonstrates a linear dependence on the drive amplitude, as shown in Fig.~\ref{fig:Rabi Oscillations}(c). Data such as these are used to calibrate the amplitude that corresponds to a $\pi$ pulse on the qubit. We repeat these measurements to calibrate the $\pi$ pulses for different gate voltages. 

We also study the Rabi oscillations as a function of drive pulse length $\sigma$ and drive frequency detuning $\delta = f_d - f_{01}$, at fixed drive amplitude. These display a characteristic chevron-like pattern as shown in Fig.~\ref{fig:Rabi Oscillations}(d). As the detuning increases, the oscillation frequency increases, while the amplitude decreases. Because we applied a higher qubit drive amplitude, we could also observe Rabi oscillations for the $f_{02}/2$ transition. It appear with a much slower Rabi rate compared to oscillations at the qubit frequency $f_{01}$ because
this is a two-photon process.

\section{Coherence times}

\begin{figure*}[!ht]  
  \centering
  \includegraphics[width=15cm]{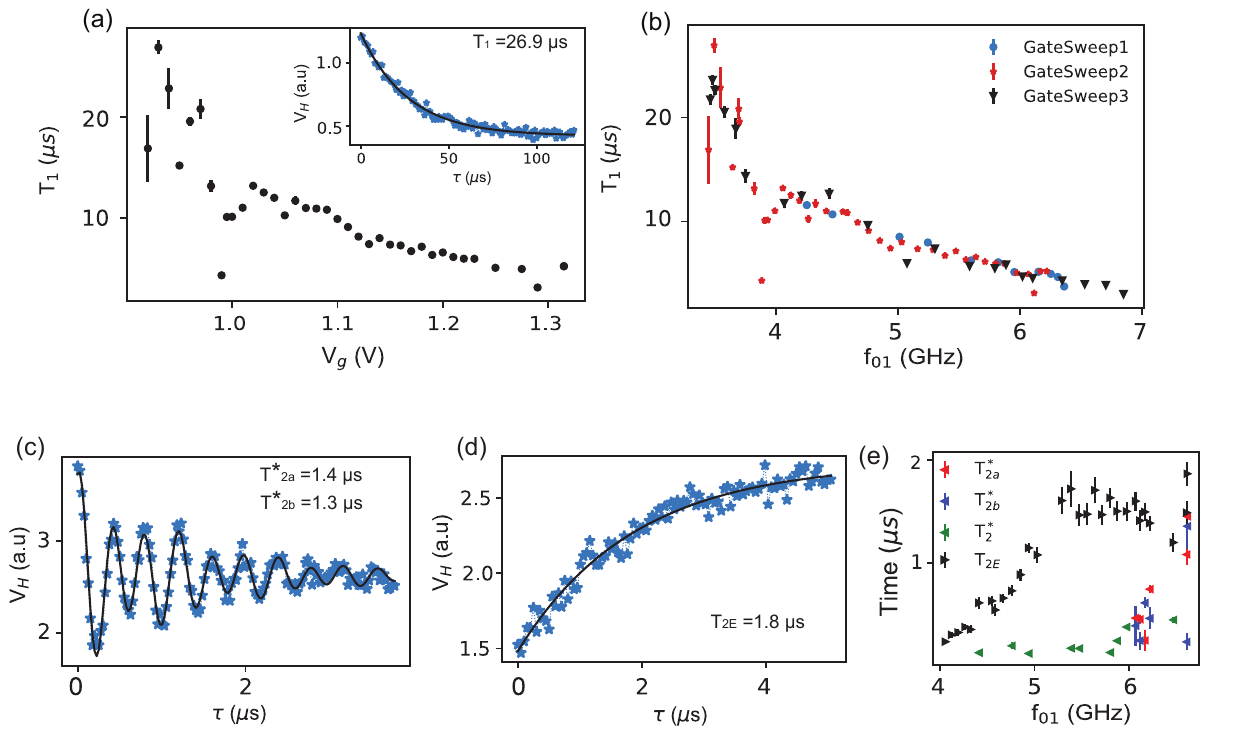}
  \caption{(a) $T_1$ as a function of side gate voltage $V_g$ for Qubit A. Inset: Relaxation time extracted by fitting an exponential to the data, 
$T_1 = 26.9 \pm 0.7\ \mu\mathrm{s}$ at $f_{01} = 3.494~\mathrm{GHz}$. 
(b) $T_1$ plotted as a function of qubit frequency $f_{01}$ for multiple gate sweeps. (c) Two dephasing times of $T_{2a} = 1.4 \pm 0.1\ \mu\mathrm{s}$ and $T_{2b} = 1.3 \pm 0.2\ \mu\mathrm{s}$ are extracted by fitting the decaying oscillations to double sinusoidal functions with different exponential envelopes. Here, $f_{01}=\text{6.616~GHz}$. (d) Echo measurement gives $T_{2E} = 1.8 \pm 0.1\ \mu\mathrm{s}$. (e) $T^*_2$, $T_{2a}$ and $T_{2b}$ and $T_{2E}$ plotted as a function of $f_{01}$. The error bars in (a), (b), and (e) represent the standard deviations of the fits.}

  \label{Coherence}
\end{figure*}

We vary the time between the drive pulse and the readout pulse to measure the qubit relaxation time $T_1$. For more negative gate voltages, as the qubit frequency decreases, $T_1$ increases as shown in Fig.~\ref{Coherence}(a). The highest $T_1 = 26.9 \pm 0.7\,\mu\text{s}$ is observed close to the lowest studied frequency $f_{01} = 3.494\,\text{GHz}$ (inset of Fig.~\ref{Coherence}(a)).

Typical for nanowire junctions, different gate voltage sweeps may result in offsets in measured Josephson energies, including hysteretic behavior. These properties originate from the semiconductor of the junction. At the same time, $T_1$ as a function of $f_{01}$ is largely gate sweep-independent, as shown in Fig.~\ref{Coherence}(b). This indicates that $T_1$ is more sensitive to dielectric losses and Purcell decay through the resonator which depend on $f_{01}$ than to any mesoscopic features of the junction such as the local charge environment which may not show a distinct trend based on $f_{01}$. We observe a drop in \(T_1\) around \(f_{01} = 3.885\,\text{GHz}\) which repeats in multiple gate sweeps. We attribute this to coupling to a two-level system (TLS). 

We perform Ramsey measurements to determine the dephasing time $T_2^*$ using a Ramsey sequence. Fig.~\ref{Coherence}(c) shows Ramsey oscillations obtained at $V_g = \text{1.32 V}$, where the qubit frequency is $f_{01}=\text{6.616~GHz}$. At this particular gate voltage, we find a beating pattern in the Ramsey oscillations with two slightly different frequencies. Fitting the oscillations to two sinusoidal functions with individual exponential decay times, we obtain two coherence times $T_{2a}^* = 1.4~\mu s$ and $T_{2b}^* = 1.3~\mu s$. We attribute the beating pattern to a nearby low frequency two-level fluctuator similar to what has been reported in \cite{zheng2024coherent}.

We also perform echo measurements by applying a \(\pi\)-pulse in the middle of the Ramsey sequence to mitigate low-frequency noise. We observe a $T_{2E}=1.8~\mu$s, which is similar to $T_2^*$ as shown in Fig.~\ref{Coherence}(d). This indicates the presence of higher-frequency noise in the system. The relaxation time at this qubit frequency is \(T_1 = 4.12\,\mu s\), thus \(T_2\) is not limited by energy relaxation. 
 
In Fig.~\ref{Coherence}(e), the measured dephasing time and echo times is summarized as a function of the qubit frequency. $T_2^*$ is extracted by fitting the decaying oscillations to a single sinusoidal function. We observe a decrease in \(T_2^*\) with decreasing qubit frequency \(f_{01}\). We attribute this behavior to reduced \(E_J/E_c\) ratios at these lower frequencies, where the qubit is more sensitive to offset charge fluctuations. At \(f_{01} = \text{3.494 GHz}\), where the highest \(T_1\) times are obtained, we are unable to measure \(T_2^*\) and \(T_{2E}\) due to short timescales.

\section{Discussion}

\begin{figure}
\centering
    \includegraphics[width=8 cm]{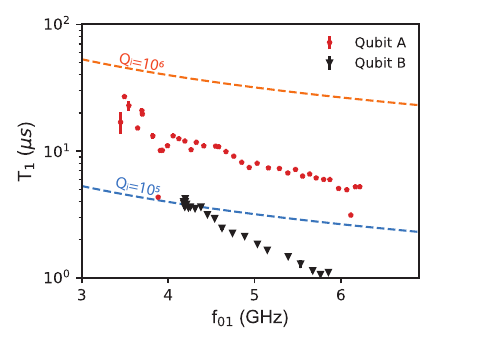}
    \caption{ $T_1$ plotted as a function of qubit frequency $f_{01}$ for Qubit A and Qubit B. The dotted lines indicate $T_1$ with constant qubit quality factor of $10^6$ and $10^5$.}

    \label{fig:LossFigure}
\end{figure}

A central question in this study is whether the longest measured coherence times are determined by the materials properties of the Sn-InAs nanowire junction, the fabrication details, or by the configuration of the external circuit.  In Fig.~\ref{fig:LossFigure}, we compare the $T_1$ data for two qubits fabricated on the same chip. The contour lines, $T_1 = Q/2\pi f_{01}$, for comparison, assume fixed quality factors. For Qubit A, the data fall close to a trend consistent with the readout resonator quality factors fabricated on the same NbTiN-Si wafer (Fig.~\ref{figApp:ResonatorQi}). This hints at the on-chip losses as the most relevant limiting factor for this qubit. Further optimization could involve varying the substrate, the groundplane superconductor and the circuit geometry.

For Qubit B, the data align closer with the lower constant quality factor dependence. This may indicate that the details of the nanowire growth and fabrication can constrain $T_1$ for some devices. We note that neither for Qubit A nor Qubit B the dependences align perfectly with the fixed quality factor lines, a deeper understanding of this would require a more elaborate modeling.

Below we discuss the various factors that may be affecting coherence times. When it comes to the nanowires, crystalline defects such as stacking faults may still be present in InAs nanowires~\cite{shtrikman2009suppression}. The angle of AlOx capping deposition is 60 $^\circ$ away from the angle of Sn deposition leads to partial capping and to the oxidation of Sn. The nanowire may land with Sn facing upwards so that the Al patch touches Sn directly, or downwards - in which case the Al patch may be contacting InAs introducing contact resistance. Ion milling may degrade the patch area introducing small contact resistance. These factors can lead to losses and limit $T_1$. At the same time, Al within the patch may act as a quasiparticle trap between two larger gap superconductors, possibly enhancing coherence.  Additionally, charge noise in the nanowires affecting the critical current may be affecting $T_2$. Operating the nanowire qubits at a voltage sweet spot, where $\partial f/\partial V_g \rightarrow 0$, could mitigate this effect. However, we do not observe clear voltage sweet spot effects. 

Factors related to the external circuit are that the qubit package box features a copper ground plane below the chip, which has been reported as limiting $T_1$ through eddy current dissipation \cite{Huang2023_EddyCurrents}. The dilution refrigerator does not include $\mu$-metal shielding, which may lead to magnetic noise. For $T_2$ times, we point out the relatively high charging energy at lower frequencies, which leads to $E_J/E_c \sim 10 - 20$  that increases the role of offset charge fluctuations~\cite{koch2007charge}.  

\section{Conclusions}

%We demonstrate coherent control of Sn-based hybrid
% superconductor-semiconductor nanowire Josephson junc-
% tion transmons.

We demonstrate coherent control of transmon qubits with Sn as the junction superconductor. The measured qubits exhibit frequency tunability of order \(3~\text{GHz}\) within a gate voltage range of \( <0.5~\text{V}\). Coherence time measurements reveal that the relaxation time \( T_1 \) scales inversely with \( f_{01} \). We measure a maximum relaxation time, \( T_1 \sim 27~\mu\text{s} \) at \( f_{01} = 3.494~\text{GHz} \) for Qubit A. The maximum $T_2$ measured is 1.8 $\mu$s at \( f_{01} = 6.616~\text{GHz} \) after an echo pulse.

As it stands, the Sn-based qubit performs on par with its Al-based hybrid qubit counterparts\cite{Luthi2018}. We identify several pathways to further enhance the qubit performance through optimization that includes lowering the charging energy, varying the substrate, ground plane and patch materials, improvements in the measurement setup, as well as further optimization of nanowire growth and fabrication. Replacing aluminum with alternative superconductors not only enables new possibilities for designing scalable quantum circuits with enhanced functionality, but also deepens our understanding of the materials aspects of the quantum device technology. \\

\section{Duration and Volume of Study} 

Qubit A and Qubit B were measured over a period of four months, during which two cooldowns were performed. The data presented in this article are from the second cooldown. In total, 20 Sn-InAs nanowire transmon devices were measured over the course of two years. Of the 20 devices measured, all exhibited spectroscopic signatures of the qubit coupled to the resonator. However, only seven devices showed measurable coherence.

\section{Data availability} 
Data and code extending beyond what is presented in the main text and the supplementary are available at \href{https://doi.org/10.5281/zenodo.16732149}{10.5281/zenodo.16732149}.

\section{Acknowledgments} 

Nanowire growth was supported by ANR HYBRID (ANR-17-PIRE-0001), ANR IMAGIQUE (ANR-42-PRC-0047), IRP HYNATOQ and the Transatlantic Research Partnership. Sn shell growth was supported by the NSF Quantum Foundry at UCSB funded via the Q-AMASE-i program under award DMR-1906325. Nanowire microscopy characterization was supported by the U.S. Department of Energy Office of Basic Energy Sciences (BES) through grant DE-SC-0019274, and transport characterization through the U.S. Department
of Energy, Basic Energy Sciences grant DE-SC-0022073. Microwave measurements were supported by the LPS/ARO nextNEQST program W911NF2210036. This work made use of the Nanoscale fabrication and Characterization facility (NFCF) at The Gertrude E. and John M. Petersen Institute of NanoScience and Engineering (PINSE) as well as of facilities at the Western Pennsylvania Quantum Information Core (WPQIC) at the University of Pittsburgh.

\bibliography{bib}

\clearpage

\renewcommand{\appendixname}{Supplementary Material}
\renewcommand{\thefigure}{S\arabic{figure}} \setcounter{figure}{0}
\renewcommand{\thetable}{S\arabic{table}} \setcounter{table}{0}
\renewcommand{\theequation}{S\arabic{table}} \setcounter{equation}{0}
\renewcommand{\thesection}{S\arabic{section}} \setcounter{section}{0}

\title{Appendix: Supporting Information and Additional Data for "Transmon qubit using Sn as a junction superconductor"}% Force line breaks with \\

\maketitle

\onecolumngrid

\section{Nanowire Growth} 

\label{App:nwGrowth}

InAs nanowires were grown using a gold-assisted vapor-liquid-solid (VLS) mechanism in a molecular beam epitaxy (MBE) reactor. Gold nanoparticles were deposited on deoxidized InAs (001) substrates, which were thermally prepared before initiating nanowire growth. The nanowires grow along the [111]B direction, resulting in inclined growth relative to the substrate surface. On (001) substrates, this leads to nanowires extending in opposing in-plane directions, producing natural crossing geometries. These geometries allow neighboring nanowires to shadow each other in later processing steps, enabling selective deposition without the need for etching.

Following nanowire growth, tin shells were deposited at cryogenic temperatures (85~K) in an ultra-high vacuum (UHV) environment. The native oxide on the nanowires was removed via atomic hydrogen cleaning prior to metal deposition. A 15~nm thin Sn layer was then evaporated at a shallow angle relative to the substrate normal, allowing in-situ shadowing to define semiconducting weak links for Josephson junctions. While the sample remained cryogenic, a 3~nm AlOx capping layer was deposited at normal incidence via electron-beam evaporation. The AlOx coating prevents Sn oxidation and de-wetting into granular structures; however, due to differing deposition angles, complete conformal coverage of the Sn shell is not warranted. Samples were subsequently warmed to room temperature under vacuum.

The nanowires so formed have diameters in the range 50-70~nm. Such etch free junctions can vary in length from 70-120~nm. It was found in transport studies that Sn induces a superconducting gap of order 600~$\mu$eV and switching currents reaching values up to 500~nA. Further details can be found in \cite{sharma2025sninasnanowireshadowdefinedjosephson}

\section{Device fabrication}
\label{App:DevFab}
The qubit devices are fabricated in multiple steps involving optical lithography and electron beam lithography. A 120~nm NbTiN film is first sputtered onto a acid (Piranha+BOE) cleaned  high resistivity ($\rho>5 k\Omega \text{cm}$) silicon wafer at 600$^{\circ}$C in AJA DC magnetron sputtering chamber. Then using optical lithography, the readout resonators, transmission line, gate line and qubit capacitors are patterned and etched with ICP-RIE chlorine etching. Fine markers with Ti/Au are defined with e-beam lithography in between the capacitor pads for positioning the Sn-InAs nanowires with a micromanipulator needle. Once positioned, SEM images are taken to locate the break in the Sn shell that defines the Josephson junction. Contacts to the Sn-InAs nanowires are designed using CAD and lithographically defined using e-beam. The connection to the rest of the superconducting circuit is established with aluminum from an e-beam evaporator after removing 3~nm of the capping $\text{AlO}_\text{x}$ layer with 80 seconds of 250V~-~15mA in-situ Argon milling. After that, the lift-off is done with acetone-IPA.

\section{Measurement Setup}
\label{App:MeasSetup}

\begin{figure}[h!]
    \centering
    \includegraphics[width=12 cm]{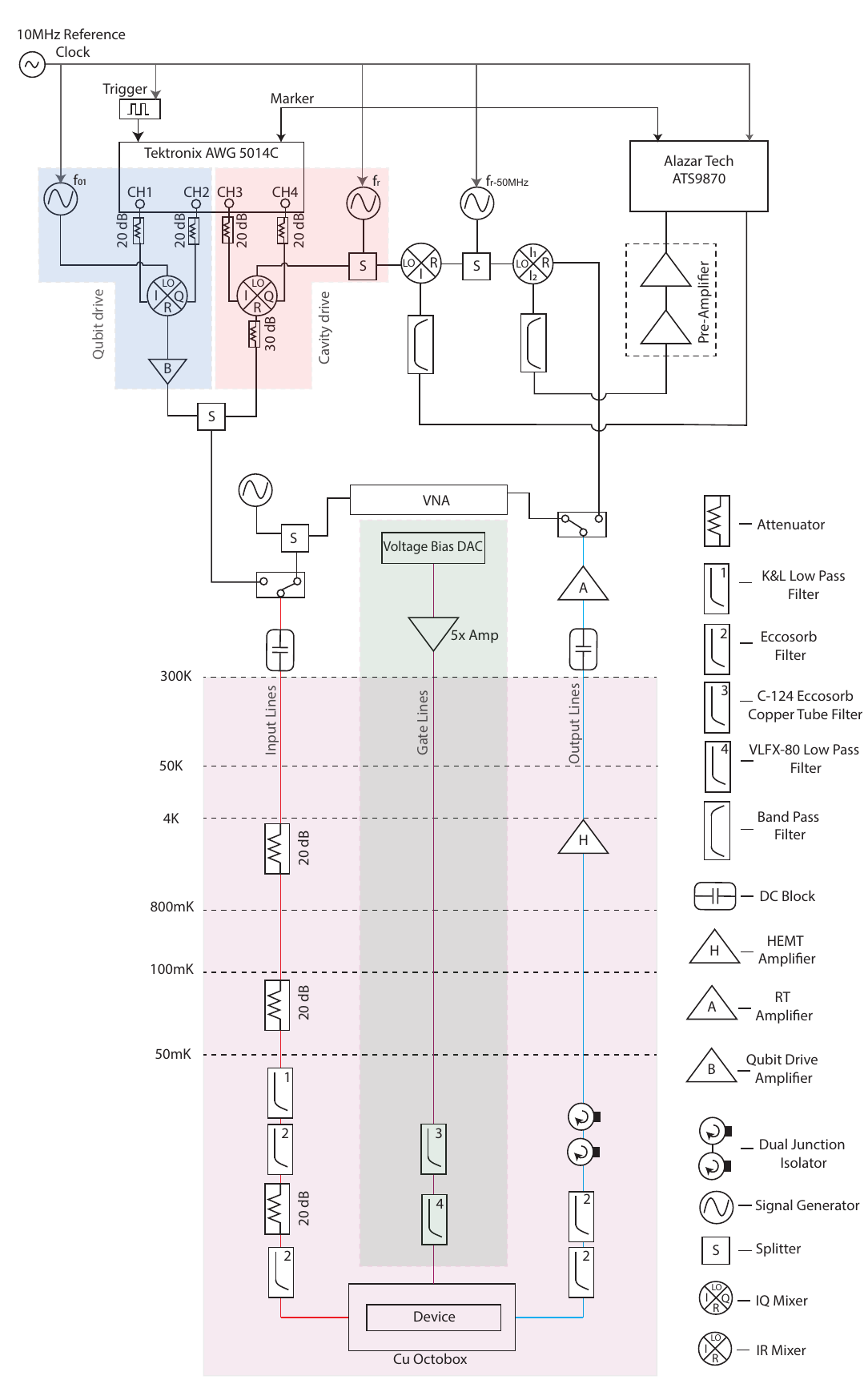}
    
    \caption{Schematic of the measurement setup along with fridge wiring. }
    \label{figApp:MeasurementSetup}
\end{figure}

All measurements discussed in the paper are performed in an Oxford triton (DR200) dilution refrigerator with a base temperature ~20mK. The schematic of the measurement setup is given in Fig.~\ref{figApp:MeasurementSetup}. 
Inside the fridge the input lines have a total attenuation of 60dB split into 20dB attenuations at different temperature stages as shown in the schematic. Homemade eccosorb filter and low pass K$\&$L filters are used at the lowest temperature stage before the input signal is routed onto the device chip. The device chip resides in a copper PCB which is placed in an copper octobox.  The output line has eccosorb filters and dual junction isolator at the MC stage. Then the signal is amplified with a HEMT amplifier at the 4K stage along with a room temperature microwave amplifier. 
The voltage to bias the gates adjacent to the semiconducting weak link of the nanowire transmons was provided by DACs of a TU Delft IVVI rack, amplified with a 5 V/V battery-driven amplifier. They are also low-pass filtered using Calmont coaxial cables along with Mini-Circuits VLFX (cutoff 80MHz) and homemade copper tube filters (filled with CR124 eccosorb-cutoff 1GHz-2dB attenuation) at the MCX before arriving at the device chip. For single tone and two tone measurements, a vector network analyzer (Keysight, E5071C) and a microwave signal generator (Keysight, N5813B) are used.
For the time domain measurements, the qubit control and readout tones are generated using SigCore5511A and modulated by a gaussian envelope with an arbitrary waveform generator (AWG, Tektronix 5014C). The output signals are demodulated and detected using AlazarTech ATS9870 digitizer with 1GS/s sampling rate. 

\section{Loss Probed by Resonator
Quality Factor}
\label{App:NbTiN_loss}

\begin{figure}[!h]
    \centering
    \includegraphics[width=9 cm]{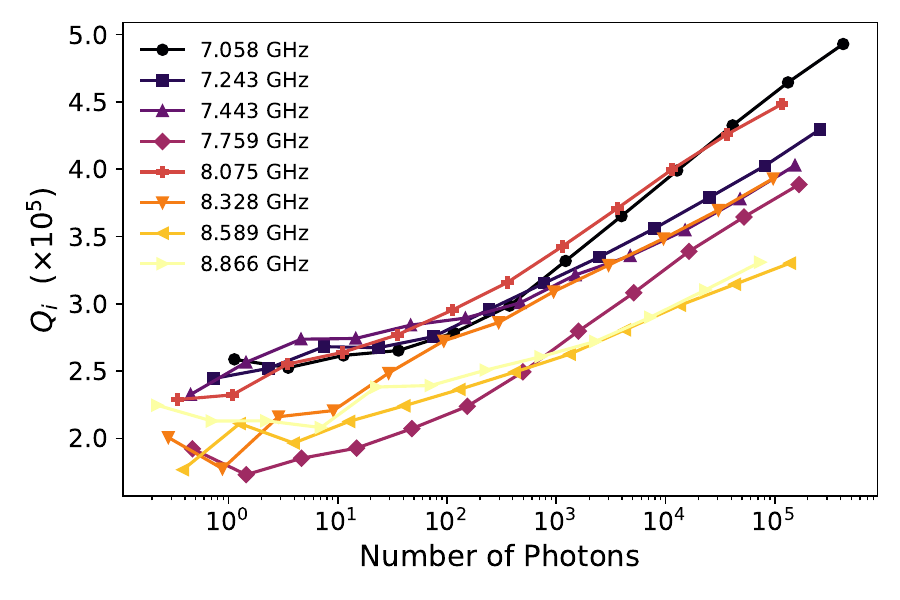}
    \caption{Power dependence of internal quality factor $Q_i$ for eight measured resonators with different resonant frequency $f_r$. All the resonators show power dependence, with $Q_i$ improving as photon number increases. This indicates $Q_i$ limited by two-level systems at low photon numbers.}
    \label{figApp:ResonatorQi}
\end{figure}

To evaluate the relative contributions of NbTiN on high resistivity silicon and the nanowire Josephson junction to qubit energy loss, we measure the quality factors of readout resonators fabricated under identical conditions but without the nanowire qubits.
We extract the $Q_i$ of these $\lambda/4$ resonators at single photon levels by functional fits of the transmission data to the following complex transmission coefficient ~\cite{Daniel_resonator_Github, khalil2012analysis, probst2015efficient}

\begin{equation}
    S_{21}=ae^{i\alpha} e^{-i 2\pi f\tau}\left[1-\dfrac{(Q_l/|Q_c|)e^{i\phi}}{1+i2Q_l(f/f_r-1)}\right]
    \label{ResonatorQi}
\end{equation}

where $a$ accounts for the net attenuation in the fridge lines, $\alpha$ is the global phase shift, $\tau$ is the electric delay, $f_r$ is the resonant frequency and $Q_c=|Q_c|e^{-i\phi}$ is the complex coupling quality factor with $\phi$ which describes the asymmetry in response due to impedance mismatches between the transmission line ports. The real valued $Q_l$ is the loaded quality factor
$
     \frac{1}{Q_l}=\frac{1}{Q_{i}}+\frac{1}{Q^{'}_{c}}
    \label{ResonatorQl}
$ where $1/Q^{'}_{c}=\frac{\cos \phi}{|Q_c|}$.
Fig.\ref{figApp:ResonatorQi} shows the plot of the fitted internal quality factors. We find that $Q_i$ increases with the input power. This behavior is explained by a power-dependent saturation effect\cite{Wang2009ResonatorTLS}, indicating that the primary source of loss in the resonators is due to surface two-level systems in Si-NbTiN.

\newpage
\section{Additional Data at low qubit frequencies for Qubit A}
\label{App:LowFreq}
\begin{figure*}[h!]
    \centering
    \includegraphics[width=15cm]{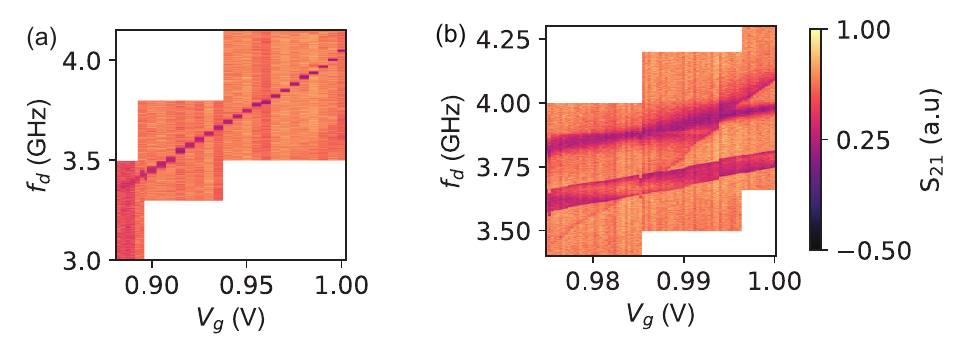}
    \caption{(a) $S_{21}$ as a function of $f_d$ and $V_g$ at low drive power near low qubit frequencies. The linewidth of $S_{21}$ dips could be seen broadening as the qubit frequency decreases with $V_g$. (b) At high drive power, the $f_{02}/2$ dip broadens and splits at certain gate voltages. A possible explanation is charge dispersion at these low qubit frequencies where $E_J/E_c\sim 15$. At around $V_g$=$\text{0.995 V}$, a state anti-crosses with the qubit at $f_{01}$ which is a TLS state where we also see a drop in $T_1$ }
    \label{S1-2}
\end{figure*}

% \subsection{}
\begin{figure*}[h!]
    \centering
    \includegraphics[width=15cm]{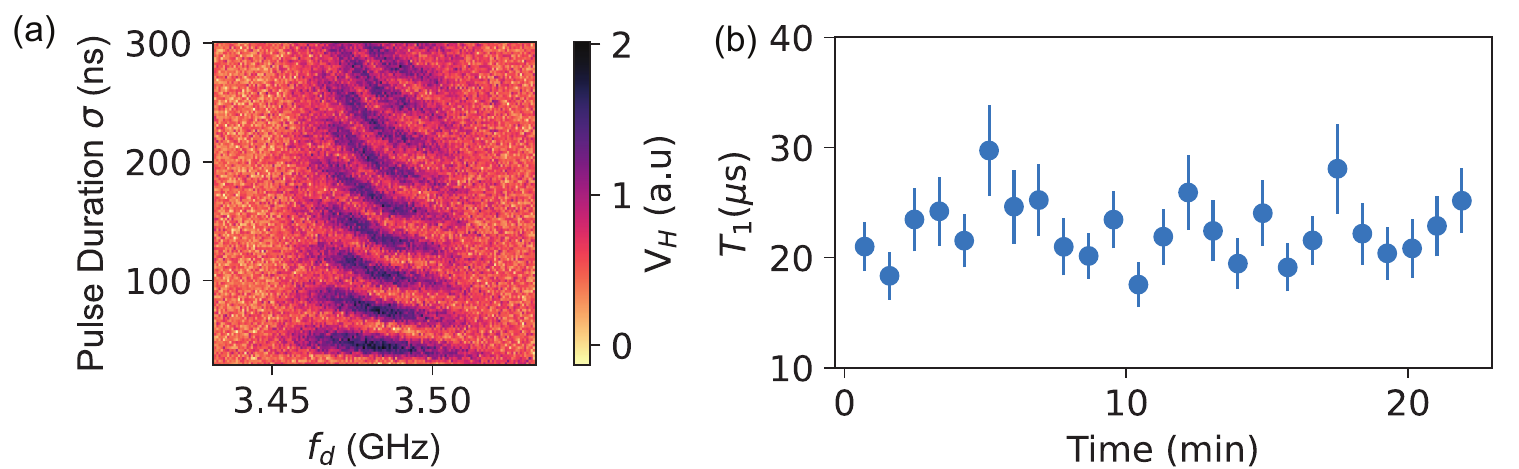}
    \caption{
   (a) Rabi oscillations showing an asymmetric chevron pattern around $f_{01} = \text{3.481~GHz}$.    
    (b) Repeated $T_1$ measurement over time for $f_{01} = \text{3.481~GHz}$ showing an average $T_1 =\text{23.6} \pm \text{0.5}~\mu\text{s}$. The error bars in (b) represent the standard deviations of the $T_1$ curve fit.}
    \label{S2-1}
\end{figure*}

\newpage
\section{Additional Data: Qubit B}
Qubit B is another device measured along side Qubit A presented in the main text. Both were designed to have the same parameters except the readout resonator frequency and went through the same fabrication process. 

\subsection{Spectroscopic measurements}
\label{QubitB-CW}
\begin{figure*}[h!]
    \centering
    \includegraphics[width=15cm]{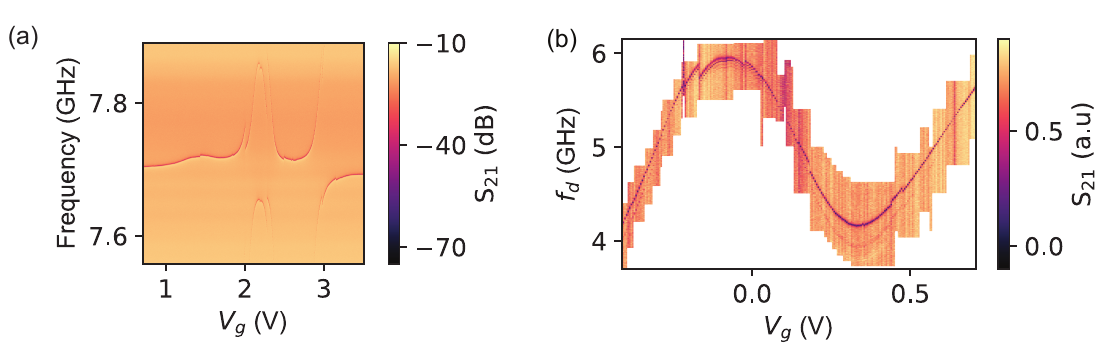}
    \caption{(a) Variation of resonator frequency $f_r$ with gate voltage $V_g$. Vacuum Rabi splittings due to qubit-resonator hybridization in gate range $1.8~\text{V} <V_g<3.2~\text{V}$. (b) Variation of \(f_{01}\) as a function of \(V_g\)  showing a tunability of \(\sim 2 \, \text{GHz}\) within a gate range of \(0.5 \, \text{V}\).}
    \label{S5-1}
\end{figure*}
\newpage
\subsection{Time Domain measurements}
\begin{figure*}[h!]
    \centering
    \includegraphics[width=15cm]{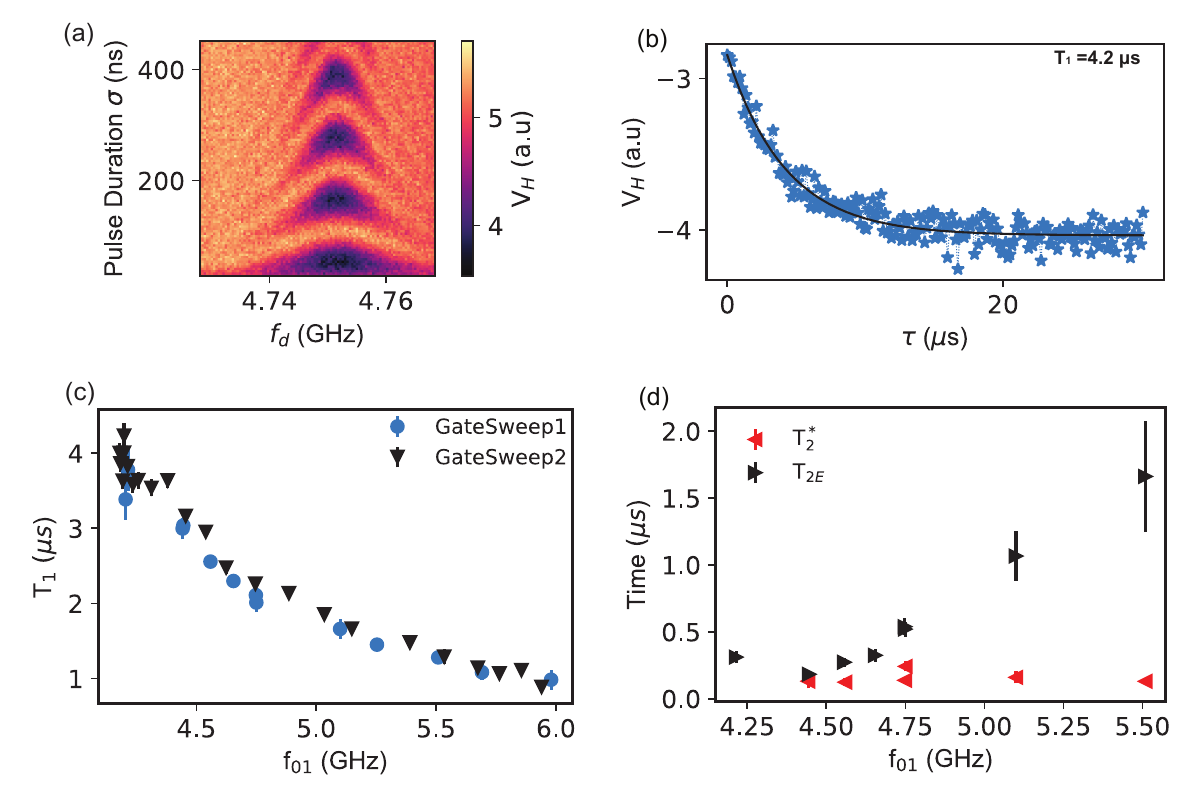}
    \caption{(a) Rabi oscillations as a function of drive frequency and pulse duration.
  (b) Decay curve fit to an exponential at \(f_{01} = 4.210 \, \text{GHz}\), yielding \(T_1 = 4.2 \pm 0.2 \, \mu\text{s}\). (c) $T_1$ as a function of $f_{01}$ for two different gate sweeps. (d) \(T_2^*\) and \(T_{2E}\) as a function of \(f_{01}\) for GateSweep1.}
    \label{S5-2}
\end{figure*}

\end{document}